\documentclass[iop]{emulateapj}

\usepackage{apjfonts}
\usepackage[T1]{fontenc}
\usepackage{color}
\usepackage{amsmath,amstext}
\usepackage{hyperref}
\usepackage{natbib}
\usepackage{graphicx}
\usepackage{gensymb}
\begin{document}

\title{Polarization constraints on the geometry of the magnetic field in the external shock of gamma-ray bursts}
\author{Eric Stringer and Davide Lazzati}
\affil{Department of Physics, Oregon State University, 301 Weniger Hall, Corvallis, OR 97331, USA}

\begin{abstract}
We study the ensemble of linear polarization measurement in the optical
afterglows of long-duration gamma-ray bursts. We assume a non
sideways-expanding top-hat jet geometry and use the relatively large
number of measurements under the assumption that they represent a
statistically unbiased sample. This allows us to constrain the ratio
between the maximum predicted polarization and the measured one, which
is an indicator of the geometry of the magnetic field in the downstream
region of the external shock. We find that the measured polarization is
substantially suppressed with respect to the maximum possible for either
a completely ordered magnetic field parallel to the shock normal or to a
field that is entirely contained in the shock plane. The measured
polarization is limited, on average, to between $25$ and $30\%$ of
the maximum theoretically possible value. This reduction requires the
perpendicular component of the magnetic field to be dominant in energy
with respect to the component parallel to the shock front, as expected
for a shock generated and/or shock compressed field. We find, however,
that the data only marginally support the assumption of a simple top-hat
jet, pointing towards a more complex geometry for the outflow.
\end{abstract}


\section{Introduction}

The structure of gamma ray-burst (GRB) jets and the origin and geometry
of their magnetic field are still openly debated. The magnetic field
within the GRB outflow is supposed to play a dominant role in the
initial collimation of the outflow, and possibly in its acceleration
(e.g., \citealt{Sasha2008}). In synchrotron models, such as the internal
shock synchrotron model \citep{Rees1994,Sari1997,Daigne1998,Burgess2019}
or the ICMART model \citep{Zhang2011} magnetic fields also play a
fundamental role in the emission of the $\gamma$-ray radiation. By
contrast, in the photospheric model radiation is advected in the outflow
and released when the flow becomes transparent
\citep{Peer2005,Peer2006,Giannios2006,Lazzati2009,Lazzati2013,
Beloborodov2011,Beloborodov2013}. Even in this case, however, magnetic
fields are expected to play an important role by providing the source of
soft photons that is required to explain the low-frequency spectrum of
typical GRBs \citep{Vurm2016}. Despite its important role, little is
known about the field's structure and origin. It could be originating
from the inner engine itself (field advected from the engine) or
generated at internal shock within the flow.

During the afterglow phase, synchrotron emission originates from the
circum-burst material swept up by the external shock and the field must
be shock-generated \citep{Meszaros1997,Sari1998}. Models to predict the
amount and temporal evolution of the linear polarization in the external
shock phase rely on asymmetries in the field structure and/or in the
outflow structure. One possibility is that the field is fully ordered in
domains or patches of a size that is smaller than the observed emitting
surface. The field among domains is, however, uncorrelated. This leads
to partial cancellation, with a residual variable linear polarization of
the order of up to $10\%$ and variable position angle
\citep{Gruzinov1999}. Alternatively, one can consider a collimated
outflow with a shock-generated field. The field is then expected to have
cylindrical symmetry around the normal to the shock front, with either
the parallel or perpendicular components dominating. If the viewer is
not perfectly aligned with the jet axis, linear polarization is
expected, with a well-defined intensity evolution and constant position
angle, except for a sudden switch of 90 degrees around the time of the
so-called jet-break \citep{Ghisellini1999,Sari1999,Granot2003}, when the
polarization momentarily vanishes. Additional variations of this
behavior are predicted in the case of structured outflows, with bright,
powerful cores surrounded by layered wings of decreasing energy
\citep{Zhang2002,Rossi2002,Granot2003b,Kumar2003}. In these geometries,
the polarization has a constant position angle and peaks around the time
of the jet break, when the powerful jet core comes into view along the
line of sight \citep{Rossi2004}.

Despite these predictions, the structure of the magnetic field remains
outstandingly unknown. The measured polarization depends on the ratio of
the magnetic energy densities in the parallel and perpendicular
components  \citep{Gruzinov1999,Sari1999}, however a firm measurement of
such ratio is hampered by the fact that the polarization depends on the
off-axis angle, which is difficult to measure robustly from the data
(e.g., \citealt{Salmonson2003,Rossi2004,vanEerten2013}). The only event
for which an off-axis angle measurement is available is the short
duration burst GRB170817A associated with the gravitational binary
merger GW170817 \citep{Lazzati2018,Mooley2018,Ghirlanda2019,Gill2018}.
In this case, however, only an upper limit for the polarization exists.
Nonetheless, a constraint on the magnetic field geometry was possible
\citep{Gill2019}.

In this paper we adopt a statistical approach by considering
observations from all available GRBs. While the off-axis angle is not
known for each individual event, one can safely assume that, within the
top-hat jet model, the probability distribution of off-axis angles
$\theta_{\rm{obs}}$ is
$p(\theta_{\rm{obs}})\propto\sin\theta_{\rm{obs}}$. In this way we can
derive a robust value for the parameter $\zeta$, the ratio of the
measured polarization to the maximum observable for a completely ordered
magnetic field. From the value of $\zeta$ we can then derive a
constraint on the magnetic field intensity ratio. As we thoroughly
discuss in Section~\ref{sec:discussion}, however, the choice of a
top-hat jet brings in some significant simplifications that cannot be
overlooked when interpreting our numerical results, and is only
marginally supported by the data.

\begin{figure}[!t]
\includegraphics[width=\columnwidth]{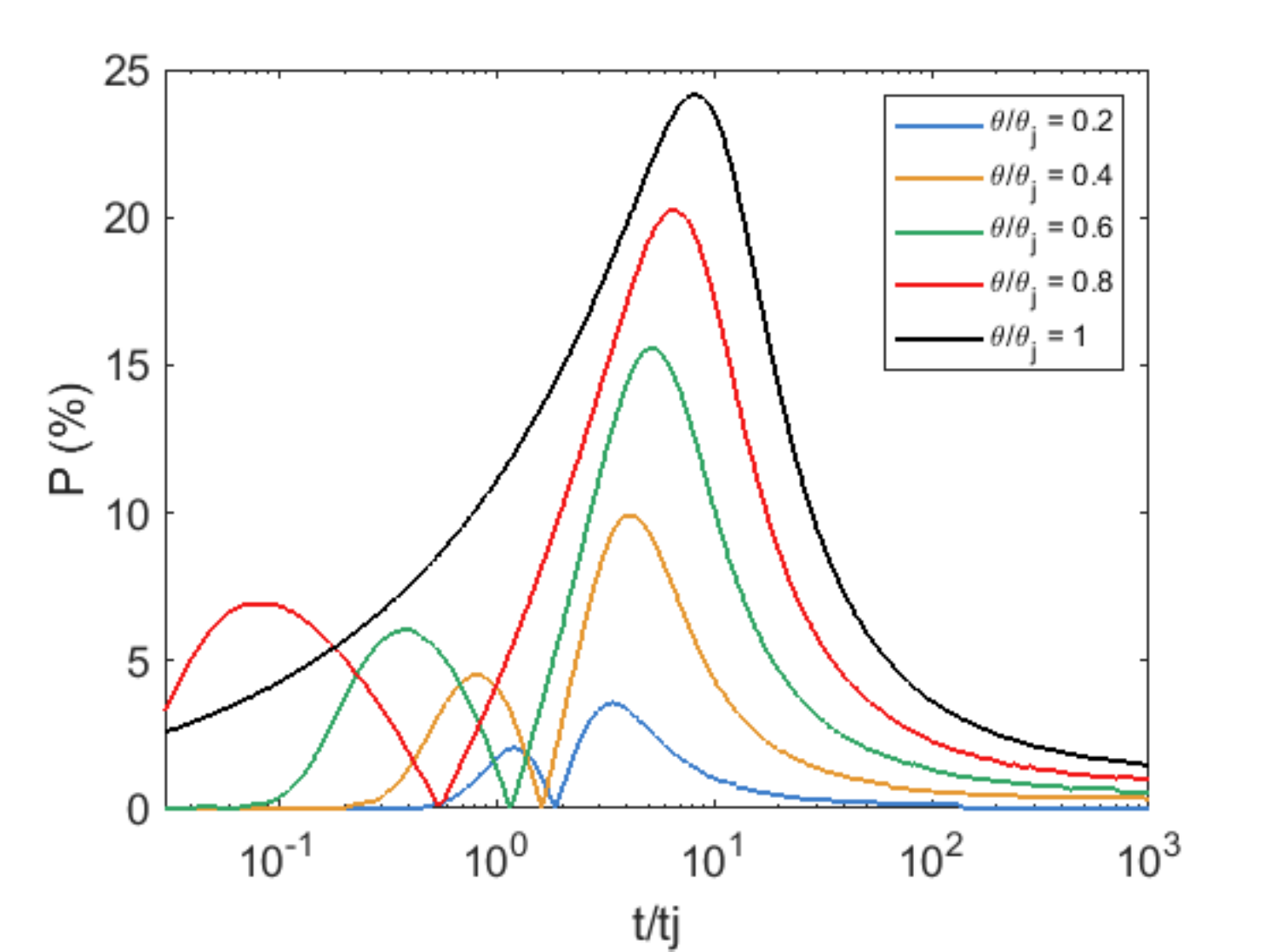}
\caption{Maximum theoretical polarization curves from a top hat jet with magnetic field entirely contained in the plane of the shock and no sideways expansion. Each line represents the polarization curve for a value of ${\theta_{obs}}/{\theta_{j}}$. Data from \cite{Rossi2004}.}
\label{fig:maxThP}
\end{figure}

\section{Methods}
In the following we describe the methodology used to derive a constraint
on the ratio between the theoretically achievable linear polarization in
a GRB afterglow ($P_{\rm{th,\,max}}$) and the measured value of linear
polarization $P$. We call this ratio $\zeta\le1$ and we therefore have:
\begin{equation}
    P=\zeta \,P_{\rm{th,\,max}}
\end{equation} 

We assume that the maximum theoretical value is given by the
calculations from various authors (e.g.,
\cite{Ghisellini1999,Sari1999,Granot2002,Granot2003,Rossi2004}) and we
adopt as our fiducial model the results from \cite{Rossi2004} for a
top-hat jet with no sideways expansion (see their Figure~8). The
resulting maximum theoretical curves are shown in
Figure~\ref{fig:maxThP} as a function of time over jet break time and
for selected values of the off-axis angle. We also assume that $\zeta$
is constant among all bursts and that it is independent of either time
or off-axis angle.

\begin{figure}[!t]
\includegraphics[width=\columnwidth]{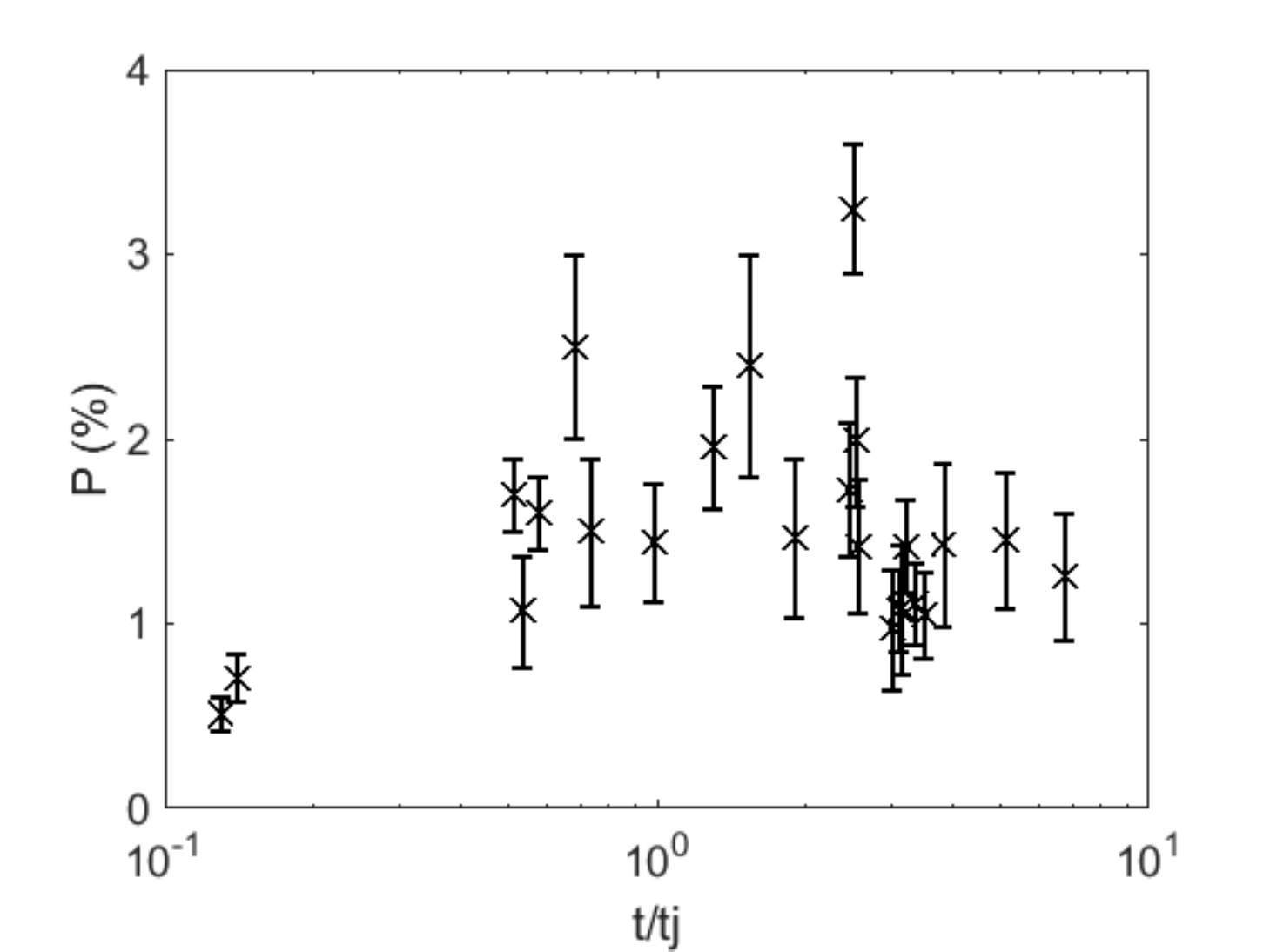}
\caption{{Linear polarization measurements used in this study versus the 
time of the measurement in units of the jet break time. Data from \cite{Covino2016}.}
\label{fig:data}}
\end{figure}

We use for our analysis the data collected in \cite{Covino2016} that
report measurements of linear polarization $P$, and the time $t$ at
which each measurement was taken with respect to the start of the burst.
An independent literature search was performed to obtain the jet break
times $t_{\rm{j}}$ for computing the measurement times in units of the
break time, as needed for the model (see Figure~\ref{fig:maxThP}). The
bursts for which reliable linear polarization and break time
measurements are available are reported in Table~\ref{tab:events}, along
with a reference for the break time estimates and polarization
measurements. Of all the data points available, there were some that did
not meet our selection criteria. Any polarization measurement with less
than $3-\sigma$ significance was discarded. In addition, any
polarization measurement with an associated jet break time of
$\frac{t}{t_j} < 0.1$ was discarded to avoid contamination from the
reverse shock emission. The burst GRB090102 was discarded because of a
large uncertainty in the jet break time \citep{Gendre2010}, while
GRB990712 was discarded due to the lack of an observable jet break
\citep{Bjornsson2001}. GRB030329 was discarded because its light curve
is complex with many re-brightening events \citep{Greiner2003}, and it
is therefore not expected to follow the model predictions. Finally,
GRB121024A was discarded since it is unique in displaying circular
polarization \citep{Wiersema2014}. The final dataset is  shown in 
Figure~\ref{fig:data}.

The main difficulty in evaluating the ratio $\zeta$ is that the
observation angle $\theta_{obs}$ for the bursts in our sample is
unknown, while in the model $\theta_{obs}$ has a dramatic effect on the
polarization levels (see Figure~\ref{fig:maxThP}). To overcome this
difficulty we use a statistical approach under the assumption that, for
a top-hat jet, the probability of detecting a burst at an angle
$\theta_{obs}$ from the jet axis is
$p(\theta_{\rm{obs}})\propto\sin\theta_{\rm{obs}}$. Since the
theoretical polarization depends on the ratio of the observer angle over
the jet opening angle $\theta_{\rm{obs}}/\theta_{\rm{j}}$, we adopt
$\theta_{\rm{j}}=10^\circ$ throughout the analysis. We will show,
however, that the final result does not depend on this particular choice
for any reasonable assumed $\theta_{\rm{j}}$.

\begin{figure}[!t]
\includegraphics[width=\columnwidth]{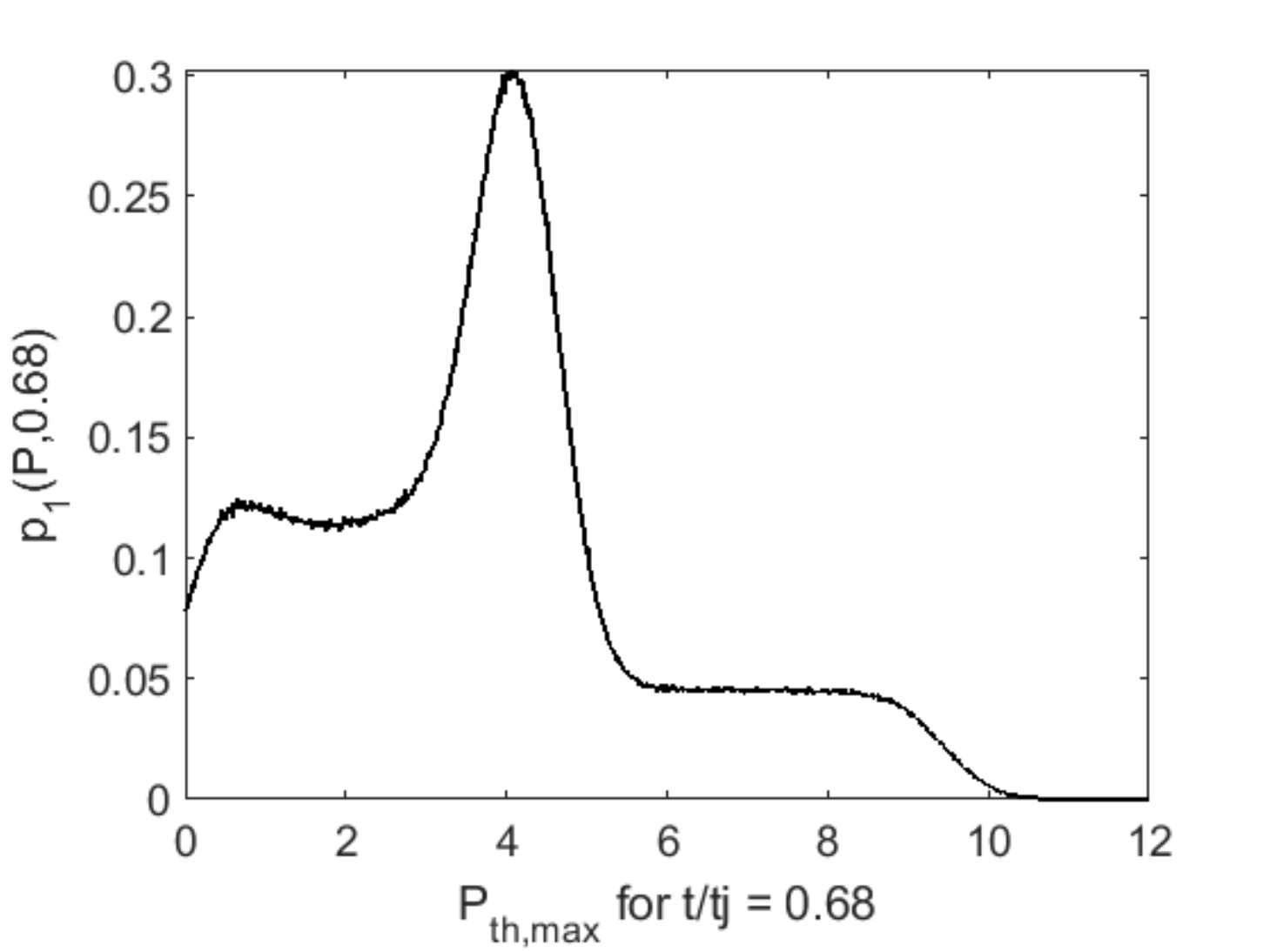}
\caption{{Polarization probability distribution for $\zeta=1$ at the time ${t_{\rm{obs}}}/{t_j}=0.68$. }
\label{fig:histogram}}
\end{figure}

For each observer time $t_{\rm{obs}}/t_{\rm{j}}$, we run a Monte Carlo
simulation drawing $10^7$ observing angles from a $\sin\theta$
probability distribution. We then use the predicted polarization
\citep{Rossi2004} to derive the probability distribution of the observed
polarization for $\zeta=1$. An example of the derived distributions is
shown in Figure~\ref{fig:histogram}. By repeating this process for every
observation we obtain a set of probability distributions, each
associated with one polarization measurement. Under our assumption that
all bursts are statistically drawn from a single population with only
the viewing angle as a variable we can now evaluate a best value for our
ratio $\zeta$ that maximizes the probability of the measurement sample.

Let  $p_1(P,t/t_{\rm{j}})\,dP$ be the probability of measuring a
polarization value within an interval $dP$ around the value $P$, for an
observed time $t/t_{\rm{j}}$ and $\zeta=1$, i.e., the maximum
theoretical value. For a given $\zeta<1$ we then have:
\begin{equation}
p_\zeta(P,t/t_{\rm{j}})=\frac{1}{\zeta} p_1\left(\frac{P}{\zeta},t/t_{\rm{j}}\right)
\end{equation}
We therefore have that the log-likelihood of all our measurements
sharing a single value of $\zeta<1$ is given by:
\begin{equation}
{\cal{L}}_{\zeta} = \displaystyle\sum_{i} \log\left\{\left[\frac{1}{\zeta}p_{1}
\left(\frac{P_i}{\zeta},t_i/t_{\rm{j}}\right)\right]\ast G(\sigma_{\rm{P_i}})\right\}
\end{equation}
where the index $i$ runs over all the measurements in
Table~\ref{tab:events}. Note that the probability distributions are
convolved with a Gaussian function $G(\sigma_{\rm{P_i}})$ with standard
deviation equal to the uncertainty of the polarization measurements.
This is done to smooth out sharp features in the theoretical probability
distributions that could not be reproduced in a dataset that is affected
by uncertainties.

\begin{table*}
\begin{center}
\begin{tabular}{||c c c c c||} 
\hline
GRB & $\frac{t}{t_j}$ & $P$ & Ref(${t_j}$) & Ref($P$)\\
\hline
 990510 & 0.51 & 1.7$\pm 0.2$ & \cite{Stanek1999} & \cite{Covino1999}\\ 
 & 0.57 & 1.6$\pm 0.2$ & & \cite{Wijers1999}\\
 020405 & 0.73 & 1.5$\pm 0.4$ & \cite{Price2003} & \cite{Masetti2003}\\ 
 & 1.3 & 1.96$\pm 0.33$ & & \cite{Covino2003}\\
 & 1.9 & 1.47$\pm 0.43$ & & \cite{Covino2003}\\
 020813 & 3.08 & 1.07$\pm 0.22$ & \cite{Lazzati2004} & \cite{Gorosabel2004}\\
 & 3.21 & 1.42$\pm 0.25$ & & \cite{Gorosabel2004}\\
 & 3.34 & 1.11$\pm 0.22$ & & \cite{Gorosabel2004}\\
 & 3.48 & 1.05$\pm 0.23$ & & \cite{Gorosabel2004}\\
 & 3.82 & 1.43$\pm 0.44$ & & \cite{Gorosabel2004}\\
 & 6.78 & 1.26$\pm 0.34$ & & \cite{Gorosabel2004}\\
 021004 & 0.13 & 0.51$\pm 0.1$ & \cite{Holland2003} & \cite{Lazzati2003}\\
 & 0.14 & 0.71$\pm 0.13$ & & \cite{Rol2003}\\
 030328 & 1.54 & 2.4$\pm 0.6$ & \cite{Maiorano2006} & \cite{Maiorano2006}\\
 080928 & 0.68 & 2.5$\pm 0.5$ & \cite{Leventis2014} & \cite{Covino2016}\\
 091018 & 0.53 & 1.07$\pm 0.3$ & \cite{Wiersema2012}& \cite{Wiersema2012}\\
 & 0.98 & 1.44$\pm 0.32$ & & \cite{Wiersema2012}\\
 & 2.46 & 1.73$\pm 0.36$ & & \cite{Wiersema2012}\\
 & 2.50 & 3.25$\pm 0.35$ & & \cite{Wiersema2012}\\
 & 2.53 & 1.99$\pm 0.35$ & & \cite{Wiersema2012}\\
 & 2.57 & 1.42$\pm 0.36$ & & \cite{Wiersema2012}\\
 & 3.01 & 0.97$\pm 0.32$ & & \cite{Wiersema2012}\\
 & 3.13 & 1.08$\pm 0.35$ & & \cite{Wiersema2012}\\
 & 5.16 & 1.45$\pm 0.37$ & & \cite{Wiersema2012}\\
\hline
\end{tabular}
\caption{The observational data used in this simulation. From left to right: Gamma ray burst, time relative to jet break time, the polarization measurement, and reference for the jet break time.}
\label{tab:events}
\end{center}
\end{table*}

\section{Results}
\label{sec:results}

\begin{figure}[!t]
\includegraphics[width=\columnwidth]{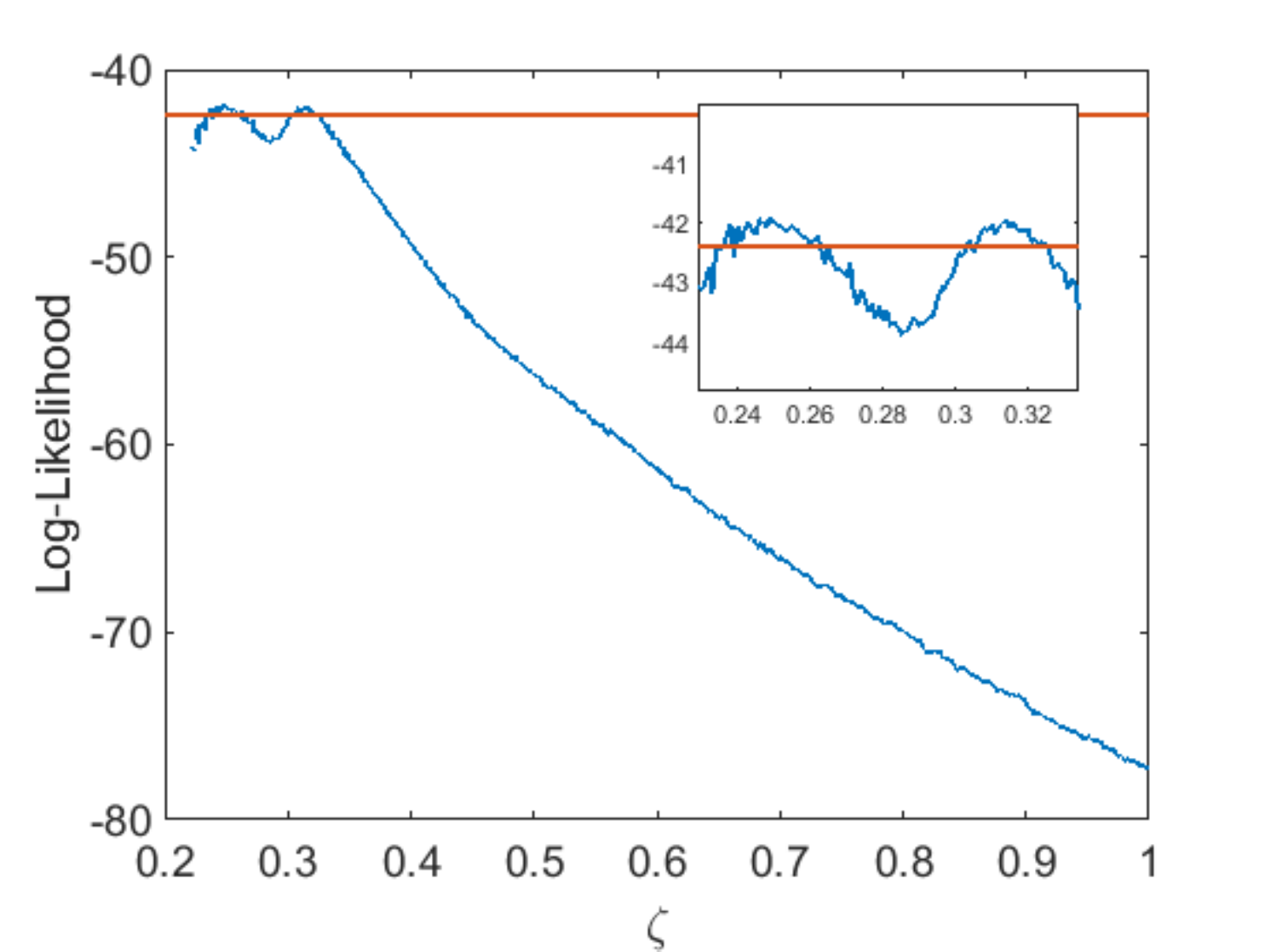}
\caption{{Log likelihood vs $\zeta$. The red line shows the
log-likelihood value of 0.5 less than the maximum that it is used to
evaluate the uncertainty on $\zeta$. The inset is a zoom of the region
around the maximum.}
\label{fig:xi}}
\end{figure}

The value of the log-likelihood as a function of the parameter $\zeta$
is shown in Figure~\ref{fig:xi}. We found that the value of $\zeta$ with
the maximum likelihood was 
\begin{equation}
    \zeta \sim 0.25
\end{equation} 
but a second high-probability peak is visible at $\zeta \sim 0.3$. The
red line in Figure~\ref{fig:xi} marks the $1-\sigma$ confidence region
and is obtained by allowing for a drop of $1/2$ in the log-likelihood
below its best value. The most likely value of $\zeta$ did not change
significantly with a change in the assumed value of $\theta_{j}$. 
The independence of the probability distribution on the assumed value of
$\theta_{j}$ also ensures that our result would remain unaltered even if
we assumed a probability distribution for the jet opening angle of
different bursts.

\begin{figure}[!t]
\includegraphics[width=\columnwidth]{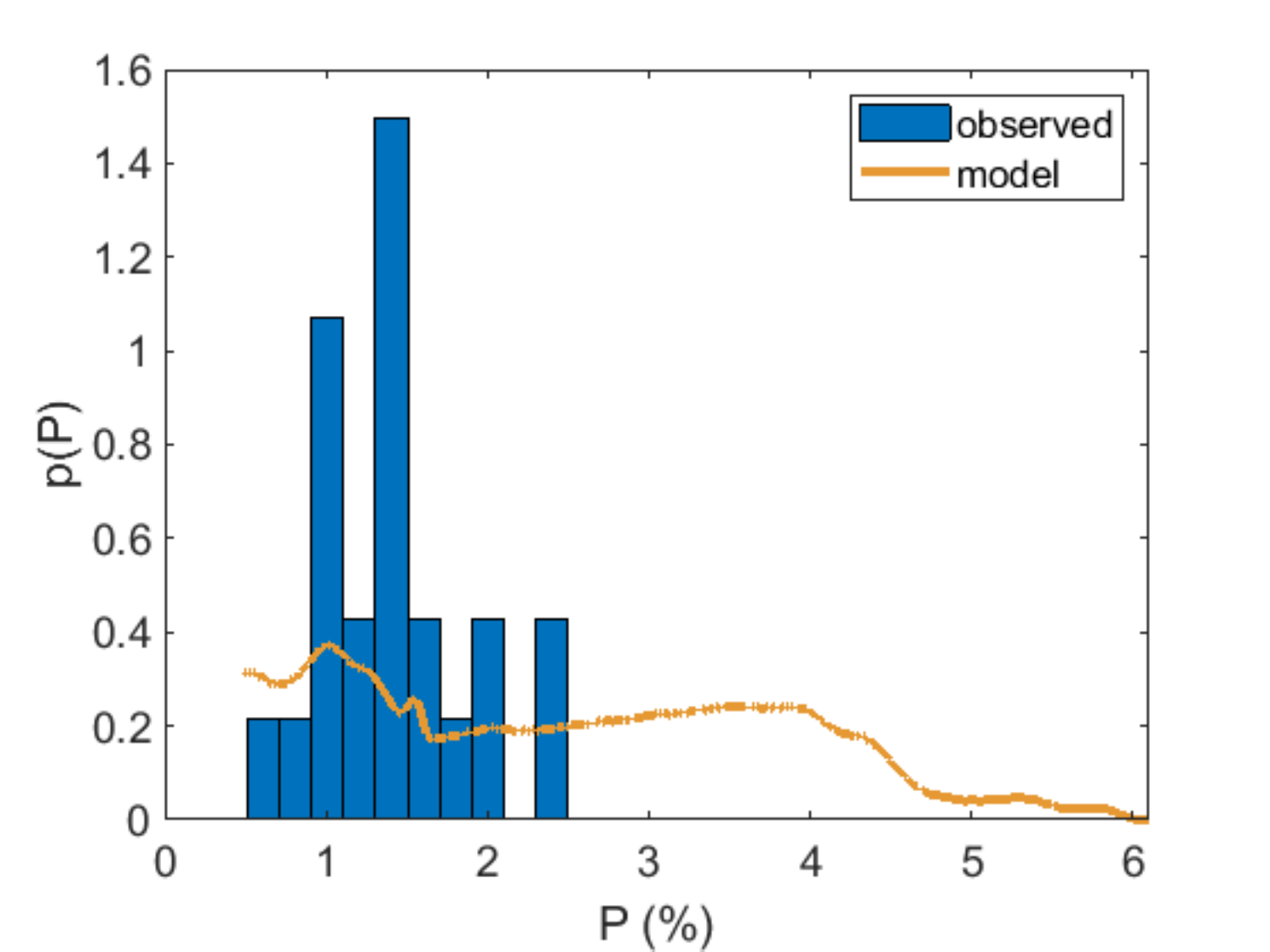}
\caption{{Comparison between the histogram of observed polarization
(blue) and the average probability distribution for the most probable
value $\zeta=0.25$ (orange solid line).}
\label{fig:KS}}
\end{figure}

The fact that $\zeta=0.25$ maximizes the likelihood does not ensure that
the model is consistent with the data. In order to verify this, we
compare the distribution of measured polarization with the average of
the individual distributions $p_{0.25}(P,t_i/t_{\rm{j}})$. The
comparison is shown in Figure~\ref{fig:KS}. A comparison of the
predicted and observed polarization rates was done using the
Kolmogorov-Smirnov test. This test resulted in a probability
$p_{\rm{KS}}=3\times10^{-5}$ that the observed and model points came
from the same distribution. As can be seen in Figure~\ref{fig:KS}, there
is a range between 2.5\% and 6\% polarization predicted in the model
that is not accounted for in the observations.

To put our result in context, consider that the observed polarization is
related to the ratio of the energy density of the parallel and
perpendicular components of the magnetic field
\citep{Gruzinov1999,Sari1999,Granot2003}

\begin{equation}
    b=2\frac{\langle B^2_\parallel \rangle}{\langle B^2_\perp \rangle}
\end{equation}
through the equation:
\begin{equation}
    \frac{P}{P_{\rm{local,\,max}}} = 
    \frac{(b-1)\sin^2\theta^\prime}{2+(b-1)\sin^2\theta^\prime}
\end{equation}
which relates the observed polarization in a given direction $P$ with
the one locally achievable in a fully ordered field
$P_{\rm{local,\,max}}$ for a viewer at a comoving angle $\theta^\prime$
from the shock normal. Since most of the radiation from a burst
afterglow comes from a thin ring at $\theta^\prime=90^\circ$
\citep{Panaitescu1998,Granot1999}, we simplify the equation as
\begin{equation}
    \frac{P}{P_{\rm{th,\,max}}} \simeq \frac{b-1}{b+1}
\end{equation}
after integration over the emitting surface. A more precise relationship
between $b$ and $\zeta$ would require a case-by-case integration over
the emitting surface that is beyond the scope of this paper (see, e.g.,
\citealt{Gill2019}). Note that the above equation predicts the
possibility of negative polarization, which is impossible. Polarization
sign, in this case, is used as an indication of the position angle of
linear polarization. Polarization of opposite sign imply position angles
orthogonal to each other. Since we cannot investigate position angle,
our value of $\zeta$ should be allowed to be negative as well as
positive. We therefore find a double constraint on the geometry of the
magnetic field:
\begin{equation}
    b=2\frac{\langle B^2_\parallel \rangle}{\langle B^2_\perp \rangle}\sim1.67;\;1.86
\end{equation}
and
\begin{equation}
    b=2\frac{\langle B^2_\parallel \rangle}{\langle B^2_\perp \rangle}\sim0.6; \;0.54
\end{equation}
where the first number corresponts to the most probable
$\zeta=0.25$ and the second to the secondary probability peak at
$\zeta=0.3$. All values  are consistent with the estimate
$0.5\lesssim{b}\lesssim2$ obtained by \cite{Granot2003}. We notice that,
in all cases, the perpendicular component of the field dominates:
$\langle B^2_\perp \rangle$ = 1.2 $\langle B^2_\parallel \rangle$, 1.1
$\langle B^2_\parallel \rangle$, 3.3 $\langle B^2_\parallel \rangle$,
and 3.7$\langle B^2_\parallel \rangle$, respectively. An alternative
parameterization was recently proposed by \cite{Gill2019}, who define
the ratio $\xi_{\rm{eff}}$ as the amount of stretch of the field along
the parallel direction with respect to an isotropic field (which would
have $\xi_{\rm{eff}}=1$). They find that $\xi_f\approx{b}^{1/2}$ and our
analysis therefore yields:
\begin{equation}
    \xi_{\rm{eff}}\sim1.3 \; ; \;\; 1.4 \; ;  \;\; 0.77 \; ;  \;\; 0.73
\end{equation}

\section{Discussion}
\label{sec:discussion}

We have presented an analysis of the available ensemble of linear
polarization measurements in the optical afterglow of long-duration
gamma-ray bursts. The analysis is aimed at finding the ratio $\zeta$
between the measured polarization and the maximum theoretically
possible, which would be observed if the magnetic field would be either
completely ordered parallel to the shock normal or entirely contained in
the plane of the shock. The ratio $\zeta$ is a proxy for the geometry of
the magnetic field that can be quantified either through the parameter
$b=2\langle B^2_\parallel \rangle/\langle B^2_\perp \rangle$
\citep{Gruzinov1999,Sari1999} or through the stretch parameter
$\xi_{\rm{eff}}$ \citep{Gill2019}.

We find that the available measurements are highly constraining of the
polarization of the afterglows and, consequently, of the field geometry.
We also find that our result is in some tension with the analysis of
\cite{Gill2019}, who looked at the upper limit on the polarization of
GW170817, a short gamma-ray burst for which the viewing geometry is
known. However, several simplifying assumptions were made and
significant systematic uncertainties are brought about by the choice of
polarization model, outflow geometry, and jet dynamics. Here we discuss
all these choices and approximations to inform the reader of their
potential impact.

First, we chose to adopt a top hat jet for our polarization model. The
rationale behind this choice is that it makes the probability of
detection of an afterglow independent on the observer angle. This makes
our calculation more straightforward and robust, since we do not have to
assume a detection threshold, nor that all bursts are observed in the
same way and with the same sensitivity. On the other hand, it seems
likely that real GRB jets are structured, at least to some level (e.g.,
\citealt{Granot2003b,Kumar2003,Rossi2004,
Zhang2004,Morsony2007,Lazzati2018}). Since the polarization curve for a
structured jet is qualitatively different from that of a top-hat jet,
our result would apply to that case only at the order of magnitude
level. We note that \cite{Gill2019} adopt a structured jet in their
analysis,  a difference that is most likely at the origin of the tension
between the two results. An extension of this work to include structured
jets is planned. As a matter of fact, the polarization measurements
shown in Figure~\ref{fig:data} do hint at a maximum of the measured
polarization at about the jet break time, a feature that is
characteristic of structured jets. Another hint that our jet model might
be oversimplified can be gleaned from Figure~\ref{fig:KS}. The figure
shows how the models predict polarization of up to $\sim6\%$, which are
not observed, yielding  a low Kolmogorov-Smirnov probability (see
Section~\ref{sec:results}). Since the highest polarization is predicted
for observers right on the edge of the jet, such prediction would be
overestimated if the jet had a smooth transition, rather than an abrupt
edge as implied by the top-hat model. In support of our choice, however,
there are at least two bursts in which the $90^\circ$ rotation of the
polarization angle that is characteristic of a top-hat jet has been
observed: GRB091018 \citep{Wiersema2012} and GRB121024A (which is,
however, excluded from the sample due to the peculiarity of having
significant circular polarization, \citealt{Wiersema2014}).

Another important choice we made was selecting the model of
\cite{Rossi2004} as our fiducial model. Calculations by different
authors differ in the details, and consequently there are differences in
the order of a few tens of percent among the theoretical models (see,
e.g., \citealt{Granot2002}). In addition, different assumptions on the
sideways expansion dynamics cause different polarization, in some cases
even adding to the complexity by causing a third peak to appear in the
polarization curve (\citealt{Sari1999}; Figure 5 in
\citealt{Rossi2004}). To test the sensitivity of our results to the
assumption on the sideways expansion, we repeated the analysis using the
models in Figure~9 of \cite{Rossi2004}, which assume that the jet
expands sideways with relativistic speed of sound $c_s=c/\sqrt{3}$. We
found a most probable value $\zeta\sim0.4$, associated with a lower KS
probability $p_{\rm{KS}}=2\times10{-5}$. These significantly different
values underline the sensitivity of polarization on poorly known dynamic
properties of the outflows. We plan to address some of these
uncertainties in future publications. Hopefully, additional data will
then be available to further constrain the models and interpretations.

\acknowledgments
We would like to thank the anonymous referee for their constructive
comments and Ramandeep Gill and Yoni Granot for reading an earlier draft
of the manuscript and making several important suggestions that
significantly improved this research. This research was supported by
NASA grants 80NSSC18K1729 (Fremi) and NNX17AK42G (ATP), Chandra grant
TM9-20002X, and NSF grant AST-1907955. 
\bibliographystyle{aasjournal}
\bibliography{biblio}

\begin{thebibliography}{}
\expandafter\ifx\csname natexlab\endcsname\relax\def\natexlab#1{#1}\fi
\providecommand{\url}[1]{\href{#1}{#1}}
\providecommand{\dodoi}[1]{doi:~\href{http://doi.org/#1}{\nolinkurl{#1}}}
\providecommand{\doeprint}[1]{\href{http://ascl.net/#1}{\nolinkurl{http://ascl.net/#1}}}
\providecommand{\doarXiv}[1]{\href{https://arxiv.org/abs/#1}{\nolinkurl{https://arxiv.org/abs/#1}}}

\bibitem[{{Beloborodov}(2011)}]{Beloborodov2011}
{Beloborodov}, A.~M. 2011, \apj, 737, 68, \dodoi{10.1088/0004-637X/737/2/68}

\bibitem[{{Beloborodov}(2013)}]{Beloborodov2013}
---. 2013, \apj, 764, 157, \dodoi{10.1088/0004-637X/764/2/157}

\bibitem[{{Bj{\"o}rnsson} {et~al.}(2001){Bj{\"o}rnsson}, {Hjorth}, {Jakobsson},
  {Christensen}, \& {Holland}}]{Bjornsson2001}
{Bj{\"o}rnsson}, G., {Hjorth}, J., {Jakobsson}, P., {Christensen}, L., \&
  {Holland}, S. 2001, \apjl, 552, L121, \dodoi{10.1086/320328}

\bibitem[{{Burgess} {et~al.}(2019){Burgess}, {B{\'e}gu{\'e}}, {Greiner},
  {Giannios}, {Bacelj}, \& {Berlato}}]{Burgess2019}
{Burgess}, J.~M., {B{\'e}gu{\'e}}, D., {Greiner}, J., {et~al.} 2019, Nature
  Astronomy, 471, \dodoi{10.1038/s41550-019-0911-z}

\bibitem[{{Covino} \& {Gotz}(2016)}]{Covino2016}
{Covino}, S., \& {Gotz}, D. 2016, Astronomical and Astrophysical Transactions,
  29, 205.
\newblock \doarXiv{1605.03588}

\bibitem[{{Covino} {et~al.}(1999){Covino}, {Lazzati}, {Ghisellini}, {Saracco},
  {Campana}, {Chincarini}, {di Serego}, {Cimatti}, {Vanzi}, {Pasquini},
  {Haardt}, {Israel}, {Stella}, \& {Vietri}}]{Covino1999}
{Covino}, S., {Lazzati}, D., {Ghisellini}, G., {et~al.} 1999, \aap, 348, L1.
\newblock \doarXiv{astro-ph/9906319}

\bibitem[{{Covino} {et~al.}(2003){Covino}, {Malesani}, {Ghisellini}, {Lazzati},
  {di Serego Alighieri}, {Stefanon}, {Cimatti}, {Della Valle}, {Fiore},
  {Goldoni}, {Kawai}, {Israel}, {Le Floc'h}, {Mirabel}, {Ricker}, {Saracco},
  {Stella}, {Tagliaferri}, \& {Zerbi}}]{Covino2003}
{Covino}, S., {Malesani}, D., {Ghisellini}, G., {et~al.} 2003, \aap, 400, L9,
  \dodoi{10.1051/0004-6361:20030133}

\bibitem[{{Daigne} \& {Mochkovitch}(1998)}]{Daigne1998}
{Daigne}, F., \& {Mochkovitch}, R. 1998, \mnras, 296, 275,
  \dodoi{10.1046/j.1365-8711.1998.01305.x}

\bibitem[{{Gendre} {et~al.}(2010){Gendre}, {Klotz}, {Palazzi}, {Kr{\"u}hler},
  {Covino}, {Afonso}, {Antonelli}, {Atteia}, {D'Avanzo}, {Bo{\"e}r}, {Greiner},
  \& {Klose}}]{Gendre2010}
{Gendre}, B., {Klotz}, A., {Palazzi}, E., {et~al.} 2010, \mnras, 405, 2372,
  \dodoi{10.1111/j.1365-2966.2010.16601.x}

\bibitem[{{Ghirlanda} {et~al.}(2019){Ghirlanda}, {Salafia}, {Paragi},
  {Giroletti}, {Yang}, {Marcote}, {Blanchard}, {Agudo}, {An}, {Bernardini},
  {Beswick}, {Branchesi}, {Campana}, {Casadio}, {Chassand e-Mottin}, {Colpi},
  {Covino}, {D'Avanzo}, {D'Elia}, {Frey}, {Gawronski}, {Ghisellini}, {Gurvits},
  {Jonker}, {van Langevelde}, {Melandri}, {Moldon}, {Nava}, {Perego},
  {Perez-Torres}, {Reynolds}, {Salvaterra}, {Tagliaferri}, {Venturi},
  {Vergani}, \& {Zhang}}]{Ghirlanda2019}
{Ghirlanda}, G., {Salafia}, O.~S., {Paragi}, Z., {et~al.} 2019, Science, 363,
  968, \dodoi{10.1126/science.aau8815}

\bibitem[{{Ghisellini} \& {Lazzati}(1999)}]{Ghisellini1999}
{Ghisellini}, G., \& {Lazzati}, D. 1999, \mnras, 309, L7,
  \dodoi{10.1046/j.1365-8711.1999.03025.x}

\bibitem[{{Giannios}(2006)}]{Giannios2006}
{Giannios}, D. 2006, \aap, 457, 763, \dodoi{10.1051/0004-6361:20065000}

\bibitem[{{Gill} \& {Granot}(2018)}]{Gill2018}
{Gill}, R., \& {Granot}, J. 2018, \mnras, 478, 4128,
  \dodoi{10.1093/mnras/sty1214}

\bibitem[{{Gill} \& {Granot}(2019)}]{Gill2019}
---. 2019, \mnras, 2980, \dodoi{10.1093/mnras/stz3340}

\bibitem[{{Gorosabel} {et~al.}(2004){Gorosabel}, {Rol}, {Covino},
  {Castro-Tirado}, {Castro Cer{\'o}n}, {Lazzati}, {Hjorth}, {Malesani}, {Della
  Valle}, {di Serego Alighieri}, {Fiore}, {Fruchter}, {Fynbo}, {Ghisellini},
  {Goldoni}, {Greiner}, {Israel}, {Kaper}, {Kawai}, {Klose}, {Kouveliotou}, {Le
  Floc'h}, {Masetti}, {Mirabel}, {M{\"o}ller}, {Ortolani}, {Palazzi}, {Pian},
  {Rhoads}, {Ricker}, {Saracco}, {Stella}, {Tagliaferri}, {Tanvir}, {van den
  Heuvel}, {Vietri}, {Vreeswijk}, {Wijers}, \& {Zerbi}}]{Gorosabel2004}
{Gorosabel}, J., {Rol}, E., {Covino}, S., {et~al.} 2004, \aap, 422, 113,
  \dodoi{10.1051/0004-6361:20034409}

\bibitem[{{Granot} \& {K{\"o}nigl}(2003)}]{Granot2003}
{Granot}, J., \& {K{\"o}nigl}, A. 2003, \apjl, 594, L83, \dodoi{10.1086/378733}

\bibitem[{{Granot} \& {Kumar}(2003)}]{Granot2003b}
{Granot}, J., \& {Kumar}, P. 2003, \apj, 591, 1086, \dodoi{10.1086/375489}

\bibitem[{{Granot} {et~al.}(2002){Granot}, {Panaitescu}, {Kumar}, \&
  {Woosley}}]{Granot2002}
{Granot}, J., {Panaitescu}, A., {Kumar}, P., \& {Woosley}, S.~E. 2002, \apjl,
  570, L61, \dodoi{10.1086/340991}

\bibitem[{{Granot} {et~al.}(1999){Granot}, {Piran}, \& {Sari}}]{Granot1999}
{Granot}, J., {Piran}, T., \& {Sari}, R. 1999, \apj, 513, 679,
  \dodoi{10.1086/306884}

\bibitem[{{Greiner} {et~al.}(2003){Greiner}, {Klose}, {Reinsch}, {Martin
  Schmid}, {Sari}, {Hartmann}, {Kouveliotou}, {Rau}, {Palazzi}, {Straubmeier},
  {Stecklum}, {Zharikov}, {Tovmassian}, {B{\"a}rnbantner}, {Ries}, {Jehin},
  {Henden}, {Kaas}, {Grav}, {Hjorth}, {Pedersen}, {Wijers}, {Kaufer}, {Park},
  {Williams}, \& {Reimer}}]{Greiner2003}
{Greiner}, J., {Klose}, S., {Reinsch}, K., {et~al.} 2003, \nat, 426, 157,
  \dodoi{10.1038/nature02077}

\bibitem[{{Gruzinov} \& {Waxman}(1999)}]{Gruzinov1999}
{Gruzinov}, A., \& {Waxman}, E. 1999, \apj, 511, 852, \dodoi{10.1086/306720}

\bibitem[{{Holland} {et~al.}(2003){Holland}, {Weidinger}, {Fynbo}, {Gorosabel},
  {Hjorth}, {Pedersen}, {M{\'e}ndez Alvarez}, {Augusteijn}, {Castro Cer{\'o}n},
  {Castro-Tirado}, {Dahle}, {Egholm}, {Jakobsson}, {Jensen}, {Levan},
  {M{\o}ller}, {Pedersen}, {Pursimo}, {Ruiz-Lapuente}, \&
  {Thomsen}}]{Holland2003}
{Holland}, S.~T., {Weidinger}, M., {Fynbo}, J. P.~U., {et~al.} 2003, \aj, 125,
  2291, \dodoi{10.1086/374235}

\bibitem[{{Kumar} \& {Granot}(2003)}]{Kumar2003}
{Kumar}, P., \& {Granot}, J. 2003, \apj, 591, 1075, \dodoi{10.1086/375186}

\bibitem[{{Lazzati} {et~al.}(2009){Lazzati}, {Morsony}, \&
  {Begelman}}]{Lazzati2009}
{Lazzati}, D., {Morsony}, B.~J., \& {Begelman}, M.~C. 2009, \apjl, 700, L47,
  \dodoi{10.1088/0004-637X/700/1/L47}

\bibitem[{{Lazzati} {et~al.}(2013){Lazzati}, {Morsony}, {Margutti}, \&
  {Begelman}}]{Lazzati2013}
{Lazzati}, D., {Morsony}, B.~J., {Margutti}, R., \& {Begelman}, M.~C. 2013,
  \apj, 765, 103, \dodoi{10.1088/0004-637X/765/2/103}

\bibitem[{{Lazzati} {et~al.}(2018){Lazzati}, {Perna}, {Morsony},
  {Lopez-Camara}, {Cantiello}, {Ciolfi}, {Giacomazzo}, \&
  {Workman}}]{Lazzati2018}
{Lazzati}, D., {Perna}, R., {Morsony}, B.~J., {et~al.} 2018, \prl, 120, 241103,
  \dodoi{10.1103/PhysRevLett.120.241103}

\bibitem[{{Lazzati} {et~al.}(2003){Lazzati}, {Covino}, {di Serego Alighieri},
  {Ghisellini}, {Vernet}, {Le Floc'h}, {Fugazza}, {Di Tomaso}, {Malesani},
  {Masetti}, {Pian}, {Oliva}, \& {Stella}}]{Lazzati2003}
{Lazzati}, D., {Covino}, S., {di Serego Alighieri}, S., {et~al.} 2003, \aap,
  410, 823, \dodoi{10.1051/0004-6361:20031321}

\bibitem[{{Lazzati} {et~al.}(2004){Lazzati}, {Covino}, {Gorosabel}, {Rossi},
  {Ghisellini}, {Rol}, {Castro Cer{\'o}n}, {Castro-Tirado}, {Della Valle}, {di
  Serego Alighieri}, {Fruchter}, {Fynbo}, {Goldoni}, {Hjorth}, {Israel},
  {Kaper}, {Kawai}, {Le Floc'h}, {Malesani}, {Masetti}, {Mazzali}, {Mirabel},
  {Moller}, {Ortolani}, {Palazzi}, {Pian}, {Rhoads}, {Ricker}, {Salmonson},
  {Stella}, {Tagliaferri}, {Tanvir}, {van den Heuvel}, {Wijers}, \&
  {Zerbi}}]{Lazzati2004}
{Lazzati}, D., {Covino}, S., {Gorosabel}, J., {et~al.} 2004, \aap, 422, 121,
  \dodoi{10.1051/0004-6361:20035951}

\bibitem[{{Leventis} {et~al.}(2014){Leventis}, {Wijers}, \& {van der
  Horst}}]{Leventis2014}
{Leventis}, K., {Wijers}, R.~A.~M.~J., \& {van der Horst}, A.~J. 2014, \mnras,
  437, 2448, \dodoi{10.1093/mnras/stt2055}

\bibitem[{{Maiorano} {et~al.}(2006){Maiorano}, {Masetti}, {Palazzi},
  {Savaglio}, {Rol}, {Vreeswijk}, {Pian}, {Price}, {Peterson}, {Jel{\'\i}nek},
  {Amati}, {Andersen}, {Castro-Tirado}, {Castro Cer{\'o}n}, {de Ugarte
  Postigo}, {Frontera}, {Fruchter}, {Fynbo}, {Gorosabel}, {Henden}, {Hjorth},
  {Jensen}, {Klose}, {Kouveliotou}, {Masi}, {M{\o}ller}, {Nicastro}, {Ofek},
  {Pand ey}, {Rhoads}, {Tanvir}, {Wijers}, \& {van den Heuvel}}]{Maiorano2006}
{Maiorano}, E., {Masetti}, N., {Palazzi}, E., {et~al.} 2006, \aap, 455, 423,
  \dodoi{10.1051/0004-6361:20054728}

\bibitem[{{Masetti} {et~al.}(2003){Masetti}, {Palazzi}, {Pian}, {Simoncelli},
  {Hunt}, {Maiorano}, {Levan}, {Christensen}, {Rol}, {Savaglio}, {Falomo},
  {Castro-Tirado}, {Hjorth}, {Delsanti}, {Pannella}, {Mohan}, {Pandey},
  {Sagar}, {Amati}, {Burud}, {Castro Cer{\'o}n}, {Frontera}, {Fruchter},
  {Fynbo}, {Gorosabel}, {Kaper}, {Klose}, {Kouveliotou}, {Nicastro},
  {Pedersen}, {Rhoads}, {Salamanca}, {Tanvir}, {Vreeswijk}, {Wijers}, \& {van
  den Heuvel}}]{Masetti2003}
{Masetti}, N., {Palazzi}, E., {Pian}, E., {et~al.} 2003, \aap, 404, 465,
  \dodoi{10.1051/0004-6361:20030491}

\bibitem[{{M{\'e}sz{\'a}ros} \& {Rees}(1997)}]{Meszaros1997}
{M{\'e}sz{\'a}ros}, P., \& {Rees}, M.~J. 1997, \apj, 476, 232,
  \dodoi{10.1086/303625}

\bibitem[{{Mooley} {et~al.}(2018){Mooley}, {Deller}, {Gottlieb}, {Nakar},
  {Hallinan}, {Bourke}, {Frail}, {Horesh}, {Corsi}, \&
  {Hotokezaka}}]{Mooley2018}
{Mooley}, K.~P., {Deller}, A.~T., {Gottlieb}, O., {et~al.} 2018, \nat, 561,
  355, \dodoi{10.1038/s41586-018-0486-3}

\bibitem[{{Morsony} {et~al.}(2007){Morsony}, {Lazzati}, \&
  {Begelman}}]{Morsony2007}
{Morsony}, B.~J., {Lazzati}, D., \& {Begelman}, M.~C. 2007, \apj, 665, 569,
  \dodoi{10.1086/519483}

\bibitem[{{Panaitescu} \& {M{\'e}sz{\'a}ros}(1998)}]{Panaitescu1998}
{Panaitescu}, A., \& {M{\'e}sz{\'a}ros}, P. 1998, \apjl, 493, L31,
  \dodoi{10.1086/311127}

\bibitem[{{Pe'er} {et~al.}(2005){Pe'er}, {M{\'e}sz{\'a}ros}, \&
  {Rees}}]{Peer2005}
{Pe'er}, A., {M{\'e}sz{\'a}ros}, P., \& {Rees}, M.~J. 2005, \apj, 635, 476,
  \dodoi{10.1086/497360}

\bibitem[{{Pe'er} {et~al.}(2006){Pe'er}, {M{\'e}sz{\'a}ros}, \&
  {Rees}}]{Peer2006}
---. 2006, \apj, 642, 995, \dodoi{10.1086/501424}

\bibitem[{{Price} {et~al.}(2003){Price}, {Kulkarni}, {Berger}, {Fox}, {Bloom},
  {Djorgovski}, {Frail}, {Galama}, {Harrison}, {McCarthy}, {Reichart}, {Sari},
  {Yost}, {Jerjen}, {Flint}, {Phillips}, {Warren}, {Axelrod}, {Chevalier},
  {Holtzman}, {Kimble}, {Schmidt}, {Wheeler}, {Frontera}, {Costa}, {Piro},
  {Hurley}, {Cline}, {Guidorzi}, {Montanari}, {Mazets}, {Golenetskii},
  {Mitrofanov}, {Anfimov}, {Kozyrev}, {Litvak}, {Sanin}, {Boynton}, {Fellows},
  {Harshman}, {Shinohara}, {Gal-Yam}, {Ofek}, \& {Lipkin}}]{Price2003}
{Price}, P.~A., {Kulkarni}, S.~R., {Berger}, E., {et~al.} 2003, \apj, 589, 838,
  \dodoi{10.1086/374730}

\bibitem[{{Rees} \& {Meszaros}(1994)}]{Rees1994}
{Rees}, M.~J., \& {Meszaros}, P. 1994, \apjl, 430, L93, \dodoi{10.1086/187446}

\bibitem[{{Rol} {et~al.}(2003){Rol}, {Wijers}, {Fynbo}, {Hjorth}, {Gorosabel},
  {Egholm}, {Castro Cer{\'o}n}, {Castro-Tirado}, {Kaper}, {Masetti}, {Palazzi},
  {Pian}, {Tanvir}, {Vreeswijk}, {Kouveliotou}, {M{\o}ller}, {Pedersen},
  {Fruchter}, {Rhoads}, {Burud}, {Salamanca}, \& {Van den Heuvel}}]{Rol2003}
{Rol}, E., {Wijers}, R.~A.~M.~J., {Fynbo}, J.~P.~U., {et~al.} 2003, \aap, 405,
  L23, \dodoi{10.1051/0004-6361:20030731}

\bibitem[{{Rossi} {et~al.}(2002){Rossi}, {Lazzati}, \& {Rees}}]{Rossi2002}
{Rossi}, E., {Lazzati}, D., \& {Rees}, M.~J. 2002, \mnras, 332, 945,
  \dodoi{10.1046/j.1365-8711.2002.05363.x}

\bibitem[{{Rossi} {et~al.}(2004){Rossi}, {Lazzati}, {Salmonson}, \&
  {Ghisellini}}]{Rossi2004}
{Rossi}, E.~M., {Lazzati}, D., {Salmonson}, J.~D., \& {Ghisellini}, G. 2004,
  \mnras, 354, 86, \dodoi{10.1111/j.1365-2966.2004.08165.x}

\bibitem[{{Salmonson}(2003)}]{Salmonson2003}
{Salmonson}, J.~D. 2003, \apj, 592, 1002, \dodoi{10.1086/375580}

\bibitem[{{Sari}(1999)}]{Sari1999}
{Sari}, R. 1999, \apjl, 524, L43, \dodoi{10.1086/312294}

\bibitem[{{Sari} \& {Piran}(1997)}]{Sari1997}
{Sari}, R., \& {Piran}, T. 1997, \mnras, 287, 110,
  \dodoi{10.1093/mnras/287.1.110}

\bibitem[{{Sari} {et~al.}(1998){Sari}, {Piran}, \& {Narayan}}]{Sari1998}
{Sari}, R., {Piran}, T., \& {Narayan}, R. 1998, \apjl, 497, L17,
  \dodoi{10.1086/311269}

\bibitem[{{Stanek} {et~al.}(1999){Stanek}, {Garnavich}, {Kaluzny}, {Pych}, \&
  {Thompson}}]{Stanek1999}
{Stanek}, K.~Z., {Garnavich}, P.~M., {Kaluzny}, J., {Pych}, W., \& {Thompson},
  I. 1999, \apjl, 522, L39, \dodoi{10.1086/312219}

\bibitem[{{Tchekhovskoy} {et~al.}(2008){Tchekhovskoy}, {McKinney}, \&
  {Narayan}}]{Sasha2008}
{Tchekhovskoy}, A., {McKinney}, J.~C., \& {Narayan}, R. 2008, \mnras, 388, 551,
  \dodoi{10.1111/j.1365-2966.2008.13425.x}

\bibitem[{{van Eerten} \& {MacFadyen}(2013)}]{vanEerten2013}
{van Eerten}, H., \& {MacFadyen}, A. 2013, \apj, 767, 141,
  \dodoi{10.1088/0004-637X/767/2/141}

\bibitem[{{Vurm} \& {Beloborodov}(2016)}]{Vurm2016}
{Vurm}, I., \& {Beloborodov}, A.~M. 2016, \apj, 831, 175,
  \dodoi{10.3847/0004-637X/831/2/175}

\bibitem[{{Wiersema} {et~al.}(2012){Wiersema}, {Curran}, {Kr{\"u}hler}, {Meland
  ri}, {Rol}, {Starling}, {Tanvir}, {van der Horst}, {Covino}, {Fynbo},
  {Goldoni}, {Gorosabel}, {Hjorth}, {Klose}, {Mundell}, {O'Brien}, {Palazzi},
  {Wijers}, {D'Elia}, {Evans}, {Filgas}, {Gomboc}, {Greiner}, {Guidorzi},
  {Kaper}, {Kobayashi}, {Kouveliotou}, {Levan}, {Rossi}, {Rowlinson}, {Steele},
  {de Ugarte Postigo}, \& {Vergani}}]{Wiersema2012}
{Wiersema}, K., {Curran}, P.~A., {Kr{\"u}hler}, T., {et~al.} 2012, \mnras, 426,
  2, \dodoi{10.1111/j.1365-2966.2012.20943.x}

\bibitem[{{Wiersema} {et~al.}(2014){Wiersema}, {Covino}, {Toma}, {van der
  Horst}, {Varela}, {Min}, {Greiner}, {Starling}, {Tanvir}, {Wijers},
  {Campana}, {Curran}, {Fan}, {Fynbo}, {Gorosabel}, {Gomboc}, {G{\"o}tz},
  {Hjorth}, {Jin}, {Kobayashi}, {Kouveliotou}, {Mundell}, {O'Brien}, {Pian},
  {Rowlinson}, {Russell}, {Salvaterra}, {di Serego Alighieri}, {Tagliaferri},
  {Vergani}, {Elliott}, {Fari{\~n}a}, {Hartoog}, {Karjalainen}, {Klose},
  {Knust}, {Levan}, {Schady}, {Sudilovsky}, \& {Willingale}}]{Wiersema2014}
{Wiersema}, K., {Covino}, S., {Toma}, K., {et~al.} 2014, \nat, 509, 201,
  \dodoi{10.1038/nature13237}

\bibitem[{{Wijers} {et~al.}(1999){Wijers}, {Vreeswijk}, {Galama}, {Rol}, {van
  Paradijs}, {Kouveliotou}, {Giblin}, {Masetti}, {Palazzi}, {Pian}, {Frontera},
  {Nicastro}, {Falomo}, {Soffitta}, \& {Piro}}]{Wijers1999}
{Wijers}, R.~A.~M.~J., {Vreeswijk}, P.~M., {Galama}, T.~J., {et~al.} 1999,
  \apjl, 523, L33, \dodoi{10.1086/312262}

\bibitem[{{Zhang} \& {M{\'e}sz{\'a}ros}(2002)}]{Zhang2002}
{Zhang}, B., \& {M{\'e}sz{\'a}ros}, P. 2002, \apj, 571, 876,
  \dodoi{10.1086/339981}

\bibitem[{{Zhang} \& {Yan}(2011)}]{Zhang2011}
{Zhang}, B., \& {Yan}, H. 2011, \apj, 726, 90,
  \dodoi{10.1088/0004-637X/726/2/90}

\bibitem[{{Zhang} {et~al.}(2004){Zhang}, {Woosley}, \& {Heger}}]{Zhang2004}
{Zhang}, W., {Woosley}, S.~E., \& {Heger}, A. 2004, \apj, 608, 365,
  \dodoi{10.1086/386300}

\end{thebibliography}

\end{document}